\documentclass[12pt]{article}
\usepackage{amsmath}
\usepackage{amsthm}
\usepackage{amsfonts}
\usepackage{color}
\usepackage{graphicx}
\usepackage{float}
\usepackage{subfig}
\usepackage{amsmath}
\usepackage{amsfonts}
\usepackage{multirow}
\usepackage{setspace}
\usepackage{array}
\usepackage{amssymb}
\usepackage{lscape}
\usepackage[authoryear]{natbib}
\usepackage{color}
\usepackage[usenames,dvipsnames]{xcolor}
\usepackage[colorlinks]{hyperref}
\usepackage{rotating}
\usepackage{tikz}
\usepackage{framed}
\usepackage{cleveref}
\usepackage{autonum}
\usepackage{bbm}
\usepackage{subfig} 

\usepackage{microtype}

\newif\ifproofs
\proofstrue   

\newif\ifblanks
\blanksfalse  

\usetikzlibrary{backgrounds}
\usetikzlibrary{arrows,shapes}
\usetikzlibrary{tikzmark}
\usetikzlibrary{calc}

\usepackage{tcolorbox}

\newcommand\blfootnote[1]{%
  \begingroup
  \renewcommand\thefootnote{}\footnote{#1}%
  \addtocounter{footnote}{-1}%
  \endgroup
}

\usepackage[margin=1in]{geometry}
\usepackage{caption}

\usepackage{attrib} 

\hypersetup{
    unicode=false,          
    pdftoolbar=true,        
    pdfmenubar=true,        
    pdffitwindow=false,     
    pdfstartview={FitH},    
    pdftitle={My title},    
    pdfauthor={Author},     
    pdfsubject={Subject},   
    pdfcreator={Creator},   
    pdfproducer={Producer}, 
    pdfkeywords={keyword1, key2, key3}, 
    pdfnewwindow=true,      
    colorlinks=true,       
    linkcolor=blue,          
    citecolor=blue,        
    filecolor=magenta,      
    urlcolor=cyan           
}

\renewcommand{\epsilon}{\varepsilon}

\newif\ifcomments
\commentstrue   

\usepackage{subfiles}

\title{Designing Simple Mechanisms}
\author{Shengwu Li\blfootnote{I thank Tilman Borgers, Eric Chen, and Jiangtao Li for valuable comments.}}
\date{{\color{red} Preliminary draft solicited for the Journal of Economic Perspectives} \\ \medskip
\today
}

\begin{document}

\maketitle

Economists today design rules for real-world marketplaces. This began in the 1990s, when economists designed auctions for radio spectrum broadcast licenses and reformed the National Resident Matching Program (NRMP), a clearinghouse that matches graduates of American medical schools to their first jobs \citep{milgrom_putting_2000,roth2002economist}.   The practice soon found myriad applications; for instance, economists designed systems to coordinate electricity generation, to match students to public schools, to swap donated goods between food banks, and to sell loans from the Bank of England to financial institutions \citep{wilson2002architecture,abdulkadirouglu2005boston,prendergast2017food,prendergast2022allocation,klemperer2010product}.

In these mechanisms, participants convey information about their preferences, which affects who gets what and (sometimes) how much they pay. Participants will naturally try to figure out whether it makes sense to provide information in accordance with their true preferences, or whether it might prove advantageous to strategize---that is whether the mechanism is incentive-compatible. If a mechanism is simple to play, then it will be easier for participants to see that it is in their interest to reveal their true preferences.

But what makes a mechanism simple? When is it easy for participants to see that a mechanism is incentive-compatible? I will start here by explaining how and why economists came to ask these questions. Then I will discuss three recent answers, which will capture different aspects of what makes a mechanism simple.

\section{Taking mechanisms literally}

Early work in mechanism design treated mechanisms as metaphors. For instance, \cite{hurwicz1973design} describes a ``resource allocation mechanism" as follows:
\begin{quotation}
    Simplifying to the utmost, we may imagine each agent having in front of him a console with one or more dials to set; the selection of dial settings by all agents determines uniquely the flow of goods and services (trade vector) between every pair of agents and also each agent's production (input-output vector), his ``trade with nature." 
\end{quotation}

Hurwicz is describing an abstract representation of richer real-world institutions. The dials stand for purchases that the agents could make, bids that they could place, bargaining strategies that they could adopt, and so on. Later work converged on a standard abstraction, in the form of the “revelation principle”: Under some conditions, it is without loss of generality to restrict attention to truthful behavior in incentive-compatible revelation mechanisms. In such mechanisms, each agent is asked to directly report their private information, and it is in their best interest to report truthfully \citep{gibbard1973manipulation, dasgupta1979implementation, myerson1979incentive}. A mechanism is \textit{dominant-strategy incentive-compatible} (`strategy-proof') if reporting truthfully is always a best response, regardless of the other players' behavior.

On the metaphorical interpretation of a mechanism, it does not matter whether participants understand that the mechanism is incentive-compatible. After all, people do not interact with the mechanism \textit{per se}; instead, the mechanism represents aspects of their everyday economic transactions. People may be proficient at arranging those transactions---making purchases in shops, haggling over contracts---and yet not recognize that the mechanism is incentive-compatible when confronted with its abstract form.

The practice of designing real-world markets led economists to focus on a more literal interpretation of mechanisms, namely that mechanism design involves ``explicitly analyzing the consequences of trading rules that presumably are really common knowledge" \citep{wilson1987}. On this view, the rules of the mechanism are not shorthand for a vague decentralized process; rather, they capture actual rules of the institution as understood by participants.  For  instance,  the rules for a sealed-bid auction specify what bids are permitted and how the winner and the payments are determined.

When we regard the mechanism as capturing the rules of a formal process, such as an auction or a school choice system, it matters whether participants understand that the mechanism is incentive-compatible. This is a pressing question for market design, for several reasons.  First, market designers often consider novel design proposals that are unfamiliar to the intended participants---or worse, may be deceptively familiar, leading participants to adopt heuristics that are unsuited to the new rules. Second, in applications such as school choice or certain high-stakes auctions, some people participate in the mechanism exactly once, and thus cannot learn from experience. Third, even if the designer asserts that the mechanism is incentive-compatible, participants may distrust that claim. Moreover, that distrust may be justified: before the intervention of economists, the National Resident Matching Program (mistakenly) claimed that its algorithm made it incentive-compatible for medical students to report their true preferences, even though this was false \citep{williams1995reexamination}.\footnote{At the time, the National Resident Matching Program ran the \textit{hospital}-proposing deferred acceptance algorithm, which is not incentive-compatible for the applicants.} Similarly, when Google started selling Internet advertising via auction, it incorrectly claimed that its auction format was strategy-proof, asserting that its “unique auction model uses  Nobel Prize-winning economic theory to eliminate [. . . ] that feeling you’ve paid too much” \citep{edelman2007internet}. (This presumably referred to the 1996 prize to William Vickrey and James Mirrlees, for work that we will shortly discuss.) To the best of my knowledge, both claims were genuine mistakes rather than intentional deception.

Participants in real-world mechanisms do not always recognize that the mechanism is incentive-compatible. The modern algorithm for the National Resident Matching Program is (for all practical purposes) incentive-compatible\footnote{The program now uses the applicant-proposing deferred acceptance algorithm, modified to account for couples. The standard deferred acceptance algorithm is strategy-proof for the proposing side \citep{dubins1981machiavelli}, but accounting for couples slightly undermines this property. Computational experiments indicate that the probability that an applicant can profitably deviate under the modern algorithm is about $1$ in $10,000$ \citep{roth1999redesign}.},  and the program prominently advises that applicants should rank jobs in their preferred order. Nonetheless, 17 percent of applicants claimed, when surveyed, to have submitted a non-truthful rank-order list \citep{rees2018suboptimal}. When \cite{rees2018experimental} recruited medical students who recently participated in the National Resident Matching Program to take part in an incentivized experiment with the same algorithm, they found that 23 percent of subjects submitted non-truthful rank-order lists in the experiment. As further evidence, in strategy-proof mechanisms that match students to degree programs, some students make unambiguous mistakes. They rank a program without a scholarship above the same program with a scholarship, even though the scholarship is worth thousands of euros. This behavior has been found in matching mechanisms for graduate psychology degrees in Israel \citep{hassidim2021limits} and for undergraduate degrees in Hungary \citep{shorrer2023dominated}.

Consequently, it is often not enough that a mechanism is theoretically incentive-compatible. Participants have to see for themselves that it is incentive-compatible; the mechanism has to be simple. Simplicity bypasses the need for participants to trust the mechanism designer. Simplicity eases the cognitive cost of participation, which is just as real as other costs. Simplicity can level the playing field, in the sense of \cite{pathak2008leveling}, by preventing unfair outcomes caused by unequal strategic sophistication. Finally, simplicity allows the designer to rely more confidently on the predictions of classical game theory, because that analysis depends on participants responding predictably and correctly to incentives.\footnote{\cite{li2024simple} make a formal critique of this last justification. If participants are strategically unsophisticated, then instead of using a simple mechanism, the designer might profitably adopt complex mechanisms that confuse the participants.}

In the rest of this article, I explain three challenges that participants may need to overcome, in order to recognize incentive compatibility: 1) thinking contingently about unobserved moves by other players; 2) planning for their own future moves; and 3) reasoning about other players’ beliefs. In each of these cases, I will explain formal criteria that capture the difficulty in question, and discuss mechanisms that alleviate that difficulty.

These three challenges are not exhaustive of the ways that mechanisms can be simple or complex. This article is not a comprehensive survey, and it is shaped by my idiosyncratic tastes and the limits of my expertise. I have omitted excellent papers for the sake of brevity.

\section{Thinking contingently about unobserved moves by other players}

Some mechanisms require participants to reason case-by-case about other players’ moves, in order to see that the mechanism is incentive-compatible.

In a seminal paper, \cite{vickrey1961counterspeculation}  invented\footnote{It is more accurate to say that Vickrey \textit{reinvented} the second-price auction. Second-price auctions were used to sell postage stamps to collectors as early as 1893 \citep{lucking2000vickrey}.} the \textit{second-price sealed-bid auction}, in which all participants simultaneously place bids, then the highest bidder wins and pays the second-highest bid.  Vickrey argued that the second-price auction is desirable because it is simple: ``Each bidder can confine his efforts and attention to an appraisal of the value the article would have in his own hands, at a considerable saving in mental strain and possibly in out-of-pocket expense."

The second-price sealed-bid auction is strategy-proof, so it is in your own best interest to bid your true value. One can see this by analogy. In a dynamic \textit{ascending auction}, the price starts low and gradually rises. At each moment, each bidder decides whether to keep bidding or to quit irrevocably. When just one bidder is left, that bidder clinches the object at the current price. It is obvious that you should keep bidding if the price is below your true value, and should quit if the price is above. Vickrey pointed out that the second-price auction is ``logically isomorphic to" the ascending auction. In the ascending auction, each bidder essentially chooses when to quit. The bidder with the highest quit-price wins, and pays the second-highest quit-price. Quit-prices in the ascending auction are equivalent to bids in the second-price auction, so it is a dominant strategy to place a bid equal to your value. 

Notice that Vickrey’s argument that the second-price sealed-bid auction is incentive-compatible works by first reasoning about a dynamic mechanism---the ascending auction---and then arguing that two mechanisms are equivalent. But do real participants treat these two auctions as equivalent?

For auctions in the wild, it can be hard to tell whether participants are bidding truthfully, because we usually do not know their value for the object. To overcome this obstacle, \cite{kagel1987information} had lab subjects bid in auctions for a virtual prize, which had a different value for each bidder. All subjects knew their own value, in dollars, for the virtual prize. The winning bidder received cash equal to their value minus the price they paid. This method enables the experimenter to observe whether subjects are bidding truthfully.

\cite{kagel1987information} randomized whether subjects participated in second-price sealed-bid auctions or in dynamic ascending auctions. Subjects played their assigned auction format 30 times in a row, with their values drawn randomly each time. Again, second-price auctions and ascending auctions are equivalent in theory; and it is a dominant strategy to choose a bid (or quit-price) equal to your value. Thus, we might expect that in both formats subjects will bid truthfully, and that the clearing price will always be equal to the second-highest value.

The theoretical prediction of truthful bidding was largely borne out in ascending auctions. Subjects rapidly figured out that truthful bidding is optimal, and in 76 percent of auctions the clearing price was at the second-highest value.

By contrast, in second-price sealed-bid auctions, subjects did not behave as theory predicted. Subjects often submitted bids far from their values, and the clearing price was at the second-highest value in no more than 20 percent of auctions. Even after 30 rounds of play, prices remained far from the theoretical prediction.

In summary, \cite{kagel1987information} found that lab subjects rapidly converge on truthful bidding in ascending auctions, but make large and persistent mistakes in second-price auctions. This result is well-replicated (for example, see \cite{kagel1993independent, mccabe1990auction,harstad2000dominant,li2017obviously,breitmoser2022obviousness}). Ascending auctions and second-price auctions are not equivalent in practice.

What did our theory miss? Why is it easy for people to see that the ascending auction is strategy-proof? Why is it hard to see this for the second-price auction?

Imagine a participant who is bidding in a second-price sealed-bid auction. Let us assume that this participant is not a trained economist, and thus does not realize that the second-price auction is ``logically isomorphic to" the ascending auction.\footnote{In a lab experiment, \cite{breitmoser2022obviousness} found that subjects are more likely to bid truthfully in second-price auctions when the mechanism is framed in a way that makes the isomorphism salient. But why does the ascending auction framing make it easier to see the dominant strategy?} Suppose that the participant values the object at \$40. To see that bidding \$40 is better than deviating to a different bid, the participant must make a case-by-case comparison, calculating payoffs for each profile of opponent bids.  In particular, the participant must understand that if a truthful bid of \$40 would win at a price of \$30 (because the highest other bid is \$30), then deviating to bid \$50 would also lead to a win at \$30. And if a truthful bid of \$40 would lose, then a bid of \$50 could only win at prices above \$40. If the participant does not keep track of each contingency, then they might be tempting to bid above their true value in order to raise the chance of winning. Of course, this strategy would be a mistake, because boosting their bid to \$50 raises their chance of winning only when the highest other bid is between \$40 and \$50. In that case, the price is so high that winning is undesirable.

Now imagine a participant who is bidding in a dynamic ascending auction, and values the object at \$40. The price is now \$30. If the participant continues to bid truthfully, then the participant will either win at a price between \$30 and \$40, or will lose and have payoff zero. Thus, every possible payoff from bidding truthfully is at least zero. On the other hand, if the participant deviates to quit right now, then their payoff is zero for sure.\footnote{We have ignored the possibility of ties. Formally, we are modeling the ascending auction as a mechanism in which we cycle between the active bidders in some fixed order, asking each to raise their bid by a dollar or to quit irrevocably. When just one bidder remains, that bidder wins and pays their last bid.} The worst-case payoff from truthful bidding is at least as good as the best-case payoff from deviating, so the participant need not keep track of different contingencies to see that truthful bidding is optimal.

A symmetric argument establishes that the participant should quit when the price exceeds their value. Suppose that the participant values  the object at \$40 and the price rises to \$41.  If the participant quits, then their payoff is zero for sure. If instead the participant keeps bidding, then their payoff can never be positive, because now they can only win at prices above \$41.

In summary, we can distinguish dynamic  ascending auctions and second-price sealed-bid auctions in this way: To play optimally in a second-price auction, you need to compare strategies case-by-case, whereas to play optimally in an ascending auction, you need not make that comparison.

In \cite{li2017obviously}, I generalized this idea to any extensive-form mechanism. A strategy is \textit{obviously dominant} if, for any deviating strategy, at any information set where the two strategies choose different actions for the first time, the worst-case payoff under the dominant strategy is at least as good as the best-case payoff under the deviation. (When the worst case from the dominant strategy is preferred to the best case from the deviation, a case-by-case comparison is not needed.) A mechanism is \textit{obviously strategy-proof} if the truthful strategy is obviously dominant.

Dyamic ascending auctions are obviously strategy-proof. In contrast, at the moment you submit your bid in a second-price sealed-bid auction, bidding your value can lead to a zero payoff (if you lose), whereas bidding above your value can yield a positive payoff (depending on the second-price bid). Thus, truthful bidding is not an obviously dominant strategy, and second-price auctions are not obviously strategy-proof.

Every obviously strategy-proof mechanism is strategy-proof. In fact, a stronger claim holds: Every obviously strategy-proof mechanism is \textit{weakly group strategy-proof}, meaning that no coalition of players can jointly deviate and all strictly benefit. We will prove the contrapositive. Suppose the mechanism is not weakly group strategy-proof. Then there exists a strategy profile and a deviating coalition, such that the deviation is strictly profitable for every member. Along the resulting path-of-play, consider the first coalition member to deviate from the truthful strategy. That player strictly benefits, so at that information set, one possible outcome from the deviating strategy is strictly better than one possible outcome from the truthful strategy. It follows that the truthful strategy is not obviously dominant, which completes the proof.

Obvious strategy-proofness characterizes the ascending auction. Suppose that we restrict attention to auctions that are efficient (allocating the object to a  bidder with maximal value) and in which only the winning bidder makes payments. In that class, an extensive-form mechanism is obviously strategy-proof if and only if it is an ascending auction \citep[p.~576--582]{NSC_2023}. 

Under weaker restrictions, one can characterize obviously strategy-proof mechanisms as `personal-clock auctions', a larger class that includes ascending auctions, descending-price reverse auctions, and some hybrid formats \citep{li2017obviously}. Roughly, in a personal-clock auction, each participant chooses between clinching a sure thing (for example,  leaving empty-handed) and a tentative alternative with a price that gets steadily worse for the participant (for example, bidding to buy the object at a price that increases each round). For another example, consider a reverse auction in which the bidder is a seller, facing a gradually descending price. At each point, the seller can either quit and keep their object (the sure thing) or offer to sell at the current price (the tentative alternative).

In a practical application of these issues, the US Federal Communications Commission auctioned off US\$19.8 billion of radio spectrum rights in 2017.  It included a reverse auction  that paid television stations to relinquish their over-the-air broadcast rights, so that this spectrum could be repurposed for wireless Internet access.

The Federal Communications Commission’s reverse auction had to deal with complicated feasibility constraints. To check whether an allocation is feasible, one must solve a graph coloring problem for a graph with up to 3,000 nodes and 2.7 million edges; the nodes represent television stations and the edges represent interference constraints. Stations that share an edge cannot be assigned to the same channel. This problem is computationally hard \citep{karp1972reducibility}. Consequentially, the reverse auction used a suite of state-of-the-art feasibility-checking algorithms to determine the allocation and payments \citep{leyton2017economics}. 

The reverse auction proceeded, roughly, as follows: The Federal Communications Commission makes an opening offer to each television station. Each television station can either quit or continue. If it quits, then it remains on the air, retaining its broadcast rights with zero net payment. If it continues, then either it sells its rights at the current price, or the FCC makes a \textit{lower} offer and the process repeats. It is an obviously dominant strategy for each television station to bid truthfully, quitting once the price drops below its value for remaining on the air. This reverse auction belongs to the class of deferred-acceptance clock auctions studied by \cite{milgrom2020clock}, which have useful implications for privacy and budget-balance.

I have not described how the reverse auction determines the offers or when it decides to close a sale. Those details depend on the complicated feasibility-checking algorithms, and most bidders lacked the computational horsepower to check whether the algorithms were running as intended. But even without those details, you can see that the reverse auction is incentive-compatible. This is no accident. A mechanism is obviously strategy-proof if and only if truth-telling can be seen to be dominant from a certain kind of partial description. Such a description specifies the sequences of queries that one participant might receive, the answers that are allowed, and how queries and answers map to possible outcomes \citep[Theorem 1]{li2017obviously}. Notice that this description omits how the outcome depends on unobserved opponent moves; in this sense, obviously strategy-proof mechanisms formalize the idea of a dominant strategy that can be recognized without contingent reasoning. 

Psychologically, it is harder to account for hypothetical contingencies, than to reason about observed events \citep{esponda2014hypothetical}. Dynamic mechanisms can help participants to avoid mistakes, by resolving uncertainty about what other players have done.\footnote{This insight connects to a larger literature on contingent reasoning \citep{martinez2019failures,cohen2022sequential,esponda2023contingent}, reviewed by \cite{niederle2023cognitive}. \cite{glazer1996extensive} formalize another sense in which dynamic games can be easier to understand than their static counterparts.}

The assumptions we make about preferences affect the structure of obviously strategy-proof mechanisms. Recent work has explored beyond auctions, characterizing obviously strategy-proof mechanisms for settings such as two-sided matching, social choice, and object allocation without transfers \citep{arribillaga_obvious_2020,arribillaga2019all,ashlagi_stable_2018,bade2019matching,bade_gibbard-satterthwaite_2017,mandal2020obviously,thomas2020classification,troyan_obviously_2019}. Other work has built tools to study obviously strategy-proof mechanisms, providing a `revelation principle' \citep{mackenzie2020revelation} and algorithms to construct them \citep{golowich2021computational}.

\section{Planning for future moves}

Some mechanisms require participants to plan far in advance for their own future moves, in order to see that the mechanism is incentive-compatible.

Suppose that we have a set of indivisible goods and a set of agents, and we want to allocate one good to each agent, without money transfers. One natural solution is to order the agents randomly, give the first agent their favorite good, then give the second agent their favorite good among those that remain, and so on iteratively. This algorithm results in an ex post efficient allocation: any other allocation that makes one agent better off, must also make another agent worse off \citep{bogomolnaia2001new}.

To carry out the above algorithm, we need to ask agents about their preferences. We could approach agents one-by-one in random order, asking each to pick their favorite good among those that remain. The resulting \textit{dynamic random priority mechanism} is obviously strategy-proof.

In contrast, we could ask each agent to submit rank-order lists of the goods, and then process those lists according to the algorithm. The resulting \textit{static random priority mechanism} is strategy-proof. However, it is not obviously strategy-proof. Because of the randomness of the order in which the rank-order lists are chosen, it is possible that if you rank goods truthfully, you might receive your third-favorite, whereas if you deviate to rank your second-favorite good at the top, then you might receive your second-favorite good, which you strictly prefer. 

In practice, dynamic random priority results in higher rates of truthful play than its static equivalent. In an incentivized laboratory experiment with four players and four goods, dynamic random priority resulted in the dominant-strategy outcome in 93 percent of games, whereas static random priority resulted in the dominant-strategy outcome in only 64 percent of games \cite[p.~3282]{li2017obviously}.\footnote{\cite{dreyfuss2022expectations} and \cite{meisner2023loss} argue that this disparity might be due to loss aversion rather than strategic mistakes.}

Ascending auctions and dynamic random priority are both intuitively simple. But some games with obviously dominant strategies are not intuitively simple. For example, take any chess position such that White can force a win, and consider the subgame that starts from that position.   If White plays the win-forcing strategy, then by definition the worst-case outcome is that White wins. If White ever deviates from that strategy, then the best-case outcome is that White wins. Thus, the win-forcing strategy is obviously dominant in the subgame. However, real chess players often fail to play the win-forcing strategy, even from positions where such a strategy has been found by computers \citep{anderson2017assessing}. (One of the unanswered questions of chess is whether white can force a win from the opening position. Chess great Bobby Fischer opined, “I think it’s almost definite that the game is a draw theoretically.”)

What makes dynamic random priority simple, and chess complicated? Intuitively, to  identify a win-forcing strategy in chess, one has to plan far in advance, consider different future contingencies, and then backward-induct to judge the merits of each current move. By contrast, when playing dynamic random priority, each player chooses exactly once, so forward planning is not required. Conventional game theory elides this distinction, because a strategy is viewed as a complete contingent plan, specifying not only what one does right now, but also what one will do at  all future contingencies.

One way that a mechanism can be simple is that it might require only limited forward planning. \cite{pycia2023theory} formalized the idea of limited forward planning. They consider games of perfect recall, that is, the information sets are such that each participant “remembers” all the past information sets that they encountered and all the past moves they took. However, at each information set, the active player can only foresee some of their other information sets. Thus, each player forms a \textit{partial strategic plan} based on their own information set, that specifies moves at foreseeable information sets. This plan is \textit{simply dominant} if the worst-case outcome from following the plan is at least as good as the best-case from deviating to another plan that chooses differently at that information set. Specifically, the worst-case and the best-case are with respect to the actions of other players and also the actions that the player takes at  unforeseeable information sets. 

This approach to modeling limited forward planning results in a variety of incentive criteria, which depend on how we specify the foreseeable information sets. At one extreme, suppose that all information sets are foreseeable; in this case, simple dominance is equivalent to obvious dominance. At the other extreme, suppose that only the present information set is foreseeable, and let us call the resulting criterion \textit{strong obvious dominance}. For example, in dynamic random priority, picking your favorite object is strongly obviously dominant. Each player is called to play exactly once, so it does not matter that future information sets are unforeseeable. Strong obvious dominance may seem too demanding a criterion, but there is a large literature studying the welfare and revenue guarantees of posted-price mechanisms, which have strongly obviously dominant strategies \citep{lucier2017economic}. 

However, one can imagine an intermediate case, where the player can foresee some, but not all, of the future information sets. Consider a bidder who values the object at \$40. The bidder is participating in an ascending auction, with the price rising by \$1 in each step. The price starts at \$1, and the bidder reasons, ``If I quit now my payoff is \$0 for sure. But if I agree to bid \$1 and plan to quit at \$2, then my payoff could be $\$40 - \$1 = \$39$,  and is at least \$0. So I'll keep going for now." Then the price rises to \$2, and the bidder thinks, ``If I quit now my payoff is \$0 for sure. But if I keep bidding, planning to quit at \$3, then my payoff could be $\$40 - \$2 = \$38$, and is at least \$0." Thus, by looking just one step ahead at each point, the bidder is led to behave in a way  that reproduces the optimal strategy. Notice that the bidder is making a collection of partial strategic plans, and these plans are not consistent with each other. When the price is \$1, they plan to quit at \$2. But when the price reaches \$2, they revise that plan.

To formalize this idea, let the foreseeable information include both the current information set (which includes all the player’s earlier moves) and every information set that is `one step ahead.' Observe that the ascending auction is \textit{one-step simple}, meaning that truthful bidding can be induced by a collection of partial strategic plans (one for each information set), each of which is simply dominant when the player looks just one step ahead. Just a little bit of foresight is enough to play optimally in an ascending auction.

Let us return to the allocation of indivisible goods without transfers. As we saw, dynamic random planning is obviously strategy-proof. But there are other obviously strategy-proof mechanisms in this setting; \cite{pycia2023theory} characterized those mechanisms, and some of them are not intuitively simple, instead requiring participants to plan far into the future in order to see that the mechanism is incentive-compatible. They showed that strengthening the simplicity requirement from obvious strategy-proofness to one-step simplicity rules out the counterintuitive mechanisms. 

It can take detailed contingent thinking to see that a static mechanism, such as the second-price sealed-bid auction, is incentive-compatible. Dynamic mechanisms, such as the ascending auction, can mitigate this difficulty by paring down the contingencies that participants must consider. But dynamic mechanisms can raise new difficulties, in some cases requiring participants to plan in advance—even far in advance—to understand that a given strategy is optimal. The concept of partial strategic plans formally introduced by \cite{pycia2023theory} provides a way  to study  this kind of complexity.\footnote{\cite{jehiel1995limited} and \cite{jehiel2007valuation} also study limited forward planning in games, but it is an open question whether one can adapt those ideas for use in mechanism design.}

\section{Reasoning about other players' beliefs}

The study of simplicity is not only about refining strategy-proofness. Even mechanisms without dominant strategies can vary in how simple they are. In particular, some mechanisms require participants to reason about other players’ beliefs, in order to see that the mechanism is incentive-compatible.

Consider bilateral trade with transfers; there is a seller who can produce an indivisible object, and a prospective buyer. The seller has a privately-known cost $C$ and the buyer has a privately-known value $V$. If a sale is made at some price $p$, then the buyer's payoff is $V - p$ and the seller's payoff is $p - C$. Otherwise, both payoffs are $0$.

One natural mechanism for bilateral trade is the double auction, studied by \cite{chatterjee1983bargaining}: The seller and buyer simultaneously submit offers, $s$ and $b$ respectively. They transact if and only if the offer $s$ from the seller is greater than or equal to the offer $b$ from the buyer. The resulting price is a weighted average of $s$ and $b$, that is, $\alpha s + (1-\alpha)b$ for parameter $\alpha \in [0,1]$. Suppose for the moment that $\alpha = 0.5$, so the mechanism splits the difference between the two offers. In this case, neither side has a dominant strategy. The seller’s utility-maximizing offer depends on the buyer’s offer $b$; if the seller knew the buyer’s offer $b$, then the seller would set $s = b$ if the seller's cost satisfies $C \leq b$, and set $s > b$ otherwise.

Suppose it is common knowledge that both players are rational, that is, they both make offers that maximize expected utility. Since both offers are made at the same time, the seller does not know the buyer's offer $b$. Instead, the seller has to form beliefs about the distribution of $b$. Since the seller knows that the buyer is rational, the seller knows that the buyer's choice of $b$ depends on the buyer's value $V$ and on the buyer's beliefs about the seller's offer $s$. And that, in turn, depends on the buyer's beliefs about the seller's cost $C$, and the buyer's beliefs about the seller's beliefs about $b$. Thus, the seller's offer depends on a second-order belief; what the seller believes that the buyer believes about $C$, as well as a third-order belief; what the seller believes that the buyer believes that the seller believes about $V$, and so on.

This many-level reasoning process seems fantastical, and in practice will be beyond the capacity of many participants. If participants are inexperienced, it seems unlikely that they will find the equilibrium of the double auction with parameter $\alpha = 0.5$ from first principles.

However, some mechanisms do not require participants to reason about higher-order beliefs. Suppose that we take the double auction, and instead set the split-the-difference parameter $\alpha$ equal to $1$.  Then the seller is effectively making a take-it-or-leave-it offer of $s$ to the buyer, and the transaction occurs at a price of $s$ if and only if $s \leq b$. In this case, it is a dominant strategy for the buyer to set $b = V$, so the buyer does not need to think about the seller's strategy. If the seller knows that the buyer will set $b = V$, then the seller’s problem reduces to choosing an offer $s$ to maximize $(s-C)$ multiplied by the probability that $s \leq V$. Observe that the solution depends on the seller’s cost $C$ and on the seller’s belief about the buyer’s value $V$, but not on any higher-order beliefs. The seller can calculate an optimal offer using only first-order beliefs, although the seller does not have a dominant strategy.

\cite{borgers_strategically_2019} proposed  a  new  incentive  criterion,  picking  out  mechanisms  in which participants can `play well' using just their first-order beliefs. The proposal is as follows: Let us fix, for each player, a set of possible utility functions (which can be thought of as the different values that each buyer might have for an object). Each player will then form a strategy, which means choosing an action based on their utility function: for example, in the double auction, the actions are the feasible offers. A strategy is \textit{undominated} if there is no other strategy that is always at least as good, and sometimes strictly better.  A \textit{first-order belief} is a distribution over the utility functions of the other players: for instance, in a double auction, the seller might have a first-order belief that the buyer’s value is uniformly distributed on all integers between 0 and 10. A strategy is \textit{robust} with respect to that first-order belief if, assuming the other players’ utilities are indeed distributed according to that belief, the strategy is a best response to every profile of undominated strategies for the other players. A mechanism is \textit{strategically simple} if for every player and every first-order belief, there exists a robust strategy.

Every strategy-proof mechanism is strategically simple, but the reverse does not hold true. For example, the double auction with split-the-difference parameter $\alpha = 1$ is strategically simple but not strategy-proof. For the buyer, offering $b = V$ is a dominant strategy, and hence a robust strategy. Moreover, offering  $b  = V$  is the buyer’s only undominated strategy, so the  seller  has  a  robust  strategy  which depends only on the seller’s cost $C$ and on the seller’s beliefs about the distribution of $V$, \textit{i.e.}\ the seller’s first-order belief. By contrast, the double auction with $\alpha = 0.5$ is not strategically simple.  The buyer has many undominated strategies, and for non-trivial seller beliefs there is no robust seller strategy. 

\cite{borgers_strategically_2019} formally proved  that  in  the  bilateral  trade  setting,  all  the  strategically simple mechanisms have a special structure: The `leading' player either chooses an offer $p$ from a set of permitted prices, or declines trade. If an offer of $p$ is made, then the `following' player either accepts (and trade occurs at price $p$) or declines, in which case no trade occurs. This class includes the double auction with $\alpha = 0$ and with $\alpha = 1$, as well as variations that restrict the set of permitted prices.

More generally, \cite{borgers_strategically_2019} found that, under a richness condition, the strategically simple mechanisms can be characterized as `local dictatorships'. Roughly, this means that if we fix a profile of utility functions, and consider the restricted mechanism that includes only actions that are undominated \textit{at that profile}, then in the restricted mechanism only one of the player's actions determines the outcome.

The meaning of what they call ``local dictatorships" is subtle, and includes mechanisms that are not ``dictatorial" in the colloquial sense. One example of a local dictatorship is a voting mechanism in which each committee member submits a list ranking all alternatives and (at the same time) the chairperson selects two alternatives for a head-to-head vote. From the selected pair of outcomes, the final outcome is the alternative that is ranked higher by a majority of the committee (suppose that all players have strict preferences over the alternatives). In this  situation, each committee member has only a single undominated action, namely, to submit their truthful rank-order list. Thus, in the restricted mechanism, the chairperson’s action determines the outcome.

\section{Conclusion}

Mechanisms can be complicated in many ways; challenging players to think contingently about other players’ moves, to plan for their own future moves, and to reason about other players’ beliefs. Thus, there are multiple ways to define what makes a mechanism “simple.”  The right criterion depends on context; after all, participants in designed mechanisms include schoolchildren, doctors, lumberjacks, fishermen, and telecommunications firms advised by game theorists \citep{athey2001information,marszalec2020epic,bulow2009winning}.

Why have formal simplicity criteria at all? After all, we know simplicity when we see it. If we use intuition to judge formal criteria, why not use intuition instead of formal criteria? I see two considerations that weigh against the intuitive approach. First, simplicity is not the only goal---there are often others, such as efficiency, fairness, or revenue. Unless we put simplicity on equal mathematical footing, we cannot study these trade-offs systematically. Without formal criteria for simplicity, we may fixate on other desiderata that are well-formalized. 

Second, our intuitive judgements about simplicity are distorted by economic training. Experts in mechanism design suffer from the curse of knowledge; it is hard to adopt the perspective of people who do not know what game theorists know \citep{camerer1989curse}. In order to produce scientific knowledge about simplicity, we must construct theories that can be tested with data.

There is a pressing need for experiments, both to test current theories about simplicity and to discover empirical regularities that future theories might explain. In testing theories, bear in mind that some tasks might be easy if done in isolation, but difficult when they are part of a broader strategic context. \cite{kagel2016auctions} survey experiments on auctions and \cite{hakimov2021experiments} survey experiments on matching mechanisms.

Much remains to be done. The criteria for simplicity discussed here are not exhaustive; there are other dimensions on which mechanisms can be simple or complex. There could be better ways to formalize the same dimensions, that are more tractable, that track human behavior more closely, or that have firmer cognitive foundations. Moreover, simplicity comes in degrees, and it would be useful to compare the relative simplicity of different mechanisms, or to find ways to say that one mechanism is “as simple as possible” given other constraints. \cite{nagel2023measure} recently made progress in this direction. 

Finally, another important direction for the simplicity literature is to study how to describe and explain mechanisms, to help participants see for themselves that the mechanism is incentive-compatible \citep{gonczarowski2023strategyproofness}. Real-world mechanisms, such as spectrum auctions and the National Resident Matching Program, often include detailed advice for participants. The Israeli Psychology Master’s Match even provided participants with a general-audience lecture demonstrating that its mechanism was strategy-proof, but some participants were clearly not convinced \citep{hassidim2021limits}. Few laboratory experiments have studied such advice, possibly due to concerns about experimenter demand effects \citep{de2019experimenter}. The experiment conducted by \cite{masuda2022net} is a rare exception.

  \bibliographystyle{ecta}
  \bibliography{references}

\end{document}